\documentclass[conference]{IEEEtran}
\usepackage{cite}
\usepackage{graphicx}
\usepackage{amsmath}  % IEEE no longer uses Computer Modern fonts, so avoid [cmex10]
\usepackage{algorithm,algpseudocode}
\usepackage{algorithmicx}
\usepackage{algpseudocode}
\usepackage{algorithm}
\usepackage{color}
\usepackage{verbatim}
\usepackage{amsfonts}
\usepackage{amsmath,amssymb}
\usepackage{url}
\usepackage{bbm}
\usepackage{bm}
\usepackage{soul}
\usepackage{stmaryrd}
\usepackage{setspace}
\usepackage{enumitem,kantlipsum}
\usepackage[compact]{titlesec}  
\usepackage[font=footnotesize]{subfig}

\DeclareMathOperator*{\maximize}{maximize}
\DeclareMathOperator{\subjectto}{subject~to}

\makeatletter
\def\endthebibliography{%
	\def\@noitemerr{\@latex@warning{Empty `thebibliography' environment}}%
	\endlist
}
\makeatother
\makeatletter
\newenvironment{varsubequations}[1]
{%
	\addtocounter{equation}{-1}%
	\begin{subequations}
		\renewcommand{\theparentequation}{#1}%
		\def\@currentlabel{#1}%
	}
	{%
	\end{subequations}
}
\makeatother

\IEEEoverridecommandlockouts
\begin{document}
	\title{Deep Reinforcement Learning for Joint Spectrum and Power Allocation in Cellular Networks}
	\author{Yasar Sinan Nasir and Dongning Guo\\
		Department of Electrical and Computer Engineering\\
		Northwestern University, Evanston, IL 60208.\\
		\thanks{This material is based upon work supported by the National Science Foundation under Grants No.~CCF-1910168 and No.~CNS-2003098 as well as a gift from Intel Incorporation.}
	}
	\maketitle
	%%%%%%%%%%%%%%%%%%%%%%%%%%%%%%%%%%%%%%%%%%%%%%%%%%%%%%%%%%%%%%%
	\begin{abstract}
		A wireless network operator typically divides the radio spectrum it possesses into a number of subbands. In a cellular network those subbands are then reused in many cells. To mitigate co-channel interference, a joint spectrum and power allocation problem is often formulated to maximize a sum-rate objective. The best known algorithms for solving such problems generally require instantaneous global channel state information and a centralized optimizer. In fact those algorithms have not been implemented in practice in large networks with time-varying subbands. Deep reinforcement learning algorithms are promising tools for solving complex resource management problems. A major challenge here is that spectrum allocation involves discrete subband selection, whereas power allocation involves continuous variables. In this paper, a learning framework is proposed to optimize both discrete and continuous decision variables. Specifically, two separate deep reinforcement learning algorithms are designed to be executed and trained simultaneously to maximize a joint objective. Simulation results show that the proposed scheme outperforms both the state-of-the-art fractional programming algorithm and a previous solution based on deep reinforcement learning.
	\end{abstract}
	\section{Introduction}
	In today's cellular networks, the spectrum is divided into many subbands. Each cellular device suffers from the co-channel interference caused by nearby access points which use the same subbands. The interference can be particularly severe with dense, irregularly placed access points. Joint subband selection and transmit power control is a crucial tool for interference mitigation.
	
	For the single band scenario, state-of-the-art optimization methods such as fractional programming (FP) \cite{shen2018fractional} have been applied to the power control problem to reach a near-optimal allocation. We assume that the number of subbands is much less than the number of cellular devices and that each link can occupy at most one subband at a time. Therefore, the joint subband selection and power allocation problem involves \emph{mixed integer programming} \cite{tan2019jointDRL}.
	
	Conventional optimization-based schemes such as fractional programming are model-driven and require a mathematically tractable and accurate model \cite{qin2019DLphysical}. Furthermore, such a scheme is in general centralized and requires instantaneous global channel state information (CSI). In addition, it reaches a solution after several iterations, and its computational complexity does not scale well for a large number of cellular devices. Therefore, its implementation is quite challenging in a practical scenario where channel conditions vary rapidly.
	
	Recently, there has been extensive research on model-free reinforcement learning based transmit power control which is purely data-driven \cite{qin2019DLphysical}. For the single band scenario, deep Q-learning has been considered on a ``centralized training and distributed execution'' framework in \cite{ghadimi2017dynamicpower, nasir2018deep, meng2018deepmulti}. Since deep Q-learning applies only to discrete power control, the continuous transmit power domain had to be quantized in \cite{ghadimi2017dynamicpower, nasir2018deep, meng2018deepmulti} which may introduce a quantization error as discussed in \cite{meng2019ddpgmulti, nasir2020DDPGpower}. Reference \cite{meng2019ddpgmulti} first showed the performance in \cite{nasir2018deep} can be improved by quantizing the transmit power using logarithmic step size instead of linear step size, and propose replacing deep Q-learning algorithm by an actor-critic learning algorithm called deep deterministic policy gradient that applies to continuous power control. 
	
	For the multiple band scenario, Tan \emph{et al.} \cite{tan2019jointDRL} have proposed to train a single deep Q-network that jointly handles both subband selection and transmit power control. One major drawback of this approach is that the action space is the Cartesian product of available subbands and quantized transmit power levels. Therefore, the deep Q-network output layer size and the number of state action pairs to be visited for convergence during training do not scale well with increasing number of subbands. Moreover, the joint deep Q-learning approach is not directly applicable to a problem that includes both discrete and continuous variables. To overcome these challenges, we propose a novel approach that consists of two layers, where the bottom layer is responsible for continuous power allocation at the physical layer by adapting deep Q-learning, and the top layer does discrete subband scheduling using deep deterministic policy gradient. Using simulations, we evaluate the proposed learning scheme by comparing it with the joint deep Q-learning approach and the fractional programming algorithm in terms of convergence rate and achieved sum-rate performance.
	\section{System Model}
	In this paper, we consider a cellular network with $N$ links that are placed in $K$ cells and share $M$ subbands. We denote the set of link and subband indexes by $\mathcal{N}=\left\{1,\dots,N\right\}$ and $\mathcal{M}=\left\{1,\dots,M\right\}$, respectively. Link $n$ is composed of receiver $n$ and its transmitter $n$. Transmitter $n$ is placed at the corresponding cell center that includes receiver $n$ within its cell boundaries. We consider a fully synchronized time slotted system with a fixed slot duration of $T$. We assume that all transmitters and receivers are equipped with a single antenna. Due to relative scarcity of available spectrum, $K$ tends to be much larger than $M$, i.e., $K \gg M$. We let each link pick one subband at the beginning of each time slot.
	
	Similar to \cite{liang2017delayedCSI}, our channel model is composed of two parts: large and small scale fading. For simplicity, we assume that the large-scale fading is same across all subbands, whereas the small-scale fading is frequency selective, i.e., different across all subbands \cite{tan2019jointDRL}. Within each subband, small-scale fading is assumed to be block-fading and flat. Let $g^{(t)}_{n \to l, m}$ denote the downlink channel gain from transmitter $n$ to receiver $l$ on subband $m$ in time slot $t$:
	\begin{align}\label{eq:subband}
	g^{(t)}_{n \to l, m} &= \beta_{n\to l} \left| h^{(t)}_{n\to l, m}\right|^2, \quad t=1,2,\dots \, ,
	\end{align}
	where $\beta_{n\to l}$ is the large-scale fading that includes path loss and log-normal shadowing, and $h^{(t)}_{n\to l, m}$ is the small-scale Rayleigh fading. We assume that the large-scale fading remains the same through many time slots. Note that in case of mobile receivers, a time index can be associated with $\beta_{n\to l}$. 
	
	We adopt Jake's fading model to describe $h^{(t)}_{n\to l, m}$ \cite{liang2017delayedCSI}. Accordingly, the small-scale fading for each channel follows a first-order complex Gauss-Markov process:
	\begin{align}\label{eq:JakesModel}
	h_{n\to l, m}^{(t)} &= \rho h_{n\to l, m}^{(t-1)} + \sqrt{1 - \rho^2} e^{(t)}_{n\to l, m},
	\end{align}
	where the correlation between two successive fading blocks $\rho=J_0(2\pi f_d T)$ with $J_0(.)$ being the zeroth-order Bessel function of the first kind depending on the maximum Doppler frequency $f_d$. Besides, $h_{n\to l, m}^{(0)}$ and the channel innovation process $e^{(1)}_{n\to l, m}, e^{(2)}_{n\to l, m}, \dots$ are independent and identically distributed circularly symmetric complex Gaussian random variables with unit variance. The cells are agnostic to the specific fading statistics a priori.
	
	We use binary variables $\alpha^{(t)}_{n,m}$ to indicate the subband selection of link $n$ in time slot $t$. If link $n$ selects subband $m$ , we have $\alpha^{(t)}_{n,m} = 1$ and $\alpha^{(t)}_{n,j} = 0$, $\forall j \neq m$. 
	We denote the transmit power of transmitter $n$ in time slot $t$ as $p^{(t)}_{n}$. The signal-to-interference-plus-noise at receiver $n$ on subband $m$ in time slot $t$ is given by
	\begin{align}\label{eq:SINRmulti}
	\gamma^{(t)}_{n,m} &= \frac{\alpha^{(t)}_{n,m} g^{(t)}_{n \to n, m} p^{(t)}_{n}}{\sum_{l \neq n}\alpha^{(t)}_{l,m} g^{(t)}_{l \to n, m} p^{(t)}_{l}+\sigma^2},
	\end{align}
	where $\sigma^2$ is the additive white Gaussian noise power spectral density at receiver $n$. Assuming normalized bandwidth, the downlink spectral efficiency achieved by link $n$ on subband $m$ during time slot $t$ is
	\begin{align}\label{eq:DynRate}
	\begin{split}
	C^{(t)}_{n,m} &= \log\left(1+\gamma_{n,m}^{(t)}\right).
	\end{split}
	\end{align}
	\section{Problem Formulation}
	Denoting subband and power vectors in time slot $t$ as $\bm{\alpha}^{(t)}=\left[ \alpha^{(t)}_{1,1}, \alpha^{(t)}_{1,2}, \dots, \alpha^{(t)}_{N,M} \right]^\intercal$ and $\bm{p}^{(t)}=\left[p^{(t)}_1,\dots,p^{(t)}_N\right]^\intercal$, respectively, we formulate the sum-rate maximization problem as \cite{Luo2008dynamicspectrum,tan2019jointDRL}:
	\begin{varsubequations}{P1}
	%\begin{subequations}
	\label{eq:P1}
	\begin{align}
	\maximize_{\bm{p}^{(t)}, \bm{\alpha}^{(t)}} & \quad \sum_{n=1}^{N}C^{(t)}_{n} \label{eq:objective}\\
	\subjectto & \quad 0 \leq p^{(t)}_n \leq P_{\textrm{max}}, \forall n \in \mathcal{N}, \label{eq:power constraint}\\
			   & \quad \alpha^{(t)}_{n,m} \in \{0,1\}, \forall n \in \mathcal{N}, \forall m \in \mathcal{M}, \\
			   & \quad \sum_{m \in \mathcal{M}} \alpha^{(t)}_{n,m}=1, \forall n \in \mathcal{N},
	\end{align}
	%\end{subequations}
	\end{varsubequations}\noindent
	where $C^{(t)}_{n} = \sum_{m=1}^{M} C^{(t)}_{n,m}$ is link $n$'s achieved spectral efficiency, and \eqref{eq:power constraint} restricts the transmit power to be nonnegative and no larger than $P_{\textrm{max}}$.
	
	Unfortunately, \eqref{eq:P1} is in general non-convex and requires \emph{mixed integer programming} to be carried out for each time slot as channel varies. Even for a given subband selection $\bm{\alpha}^{(t)}$, this problem has been proven to be NP-hard \cite{Luo2008dynamicspectrum}. Conventional algorithms such as fractional programming are centralized solutions to \eqref{eq:P1}, but these algorithms still require many iterations to converge and their computational complexity does not scale well with increasing number of links. Besides that, obtaining instantaneous global CSI in a centralized controller and sending the allocation decisions back to the transmitters is quite challenging in practice.% As an alternative, we propose a deep reinforcement learning approach that can be distributively executed by local CSI. 
	
	\section{A Deep Reinforcement Learning Framework}
	\subsection{Overview of Reinforcement Learning}\label{sec:RLdiscussion}
	Model-free reinforcement learning \cite{sutton2018reinforcement} is a trial-and-error process where an agent interacts with an unknown environment in a sequence of discrete time steps to achieve a task. At time $t$, agent first observes the current state of the environment which is a tuple of relevant environment features and is denoted as $s^{(t)} \in \mathcal{S}$, where $\mathcal{S}$ is the set of possible states. It then takes an action $a^{(t)} \in \mathcal{A}$ from an allowed set of actions $\mathcal{A}$ according to a policy which can be either stochastic, i.e., $\pi$ with $a^{(t)} \sim \pi(\cdot | s^{(t)})$ or deterministic, i.e., $\mu$ with $a^{(t)} = \mu(s^{(t)})$ \cite{achiam2018spinup}. Since the interactions are often modeled as a Markov decision process, the environment moves to a next state $s^{(t+1)}$ following an unknown transition matrix that maps state-action pairs onto a distribution of next states, and the agent receives a reward $s^{(t+1)}$. Overall, the above process is described as an experience at $t+1$ denoted as $e^{(t+1)}=\left(s^{(t)},a^{(t)},r^{(t+1)},s^{(t+1)}\right)$. The goal is to learn a policy that maximizes the cumulative discounted reward at time $t$, defined as
	\begin{align}\label{eq:discountedreward}
	\begin{split}
	R^{(t)} &= \sum_{\tau=0}^{\infty}\gamma^{\tau} r^{(t+\tau+1)}
	\end{split},
	\end{align}
	where $\gamma \in (0,1]$ is the discount factor. 
	
	Next, we introduce two reinforcement learning methods that are used in the proposed design.
	
	Q-learning \cite{sutton2018reinforcement} is a popular reinforcement learning method that learns an action value function $Q(s,a)$. Let $\pi(a|s)$ be the probability of taking action $a$ conditioned on the current state being $s$. Assuming a stationary setting, the Q-function under a $\pi$ is the expected cumulative discounted reward when action $a$ is taken in state $s$:
	\begin{align}\label{eq:qfunction}
	Q^{\pi}(s,a) &= \mathbb{E}_{\pi}\left[R^{(t)} \middle| s^{(t)}=s, a^{(t)}=a \right].
	\end{align}
	Assuming the optimal policy $\pi^*(a|s)$ be equal to 1 for the most favorable action $a^*$ that maximizes $Q^{\pi^*}(s,a)$ for a given state $s$, the optimal Q-function satisfies the Bellman equation:
	\begin{align}\label{eq:bellmaneq}
	Q^{\pi^*}(s,a)=\mathcal{R}(s,a)+\gamma \sum_{s' \in S}\mathcal{P}^{a}_{ss'}\max_{a'}Q^{\pi^*}(s',a'),
	\end{align}
	where $\mathcal{R}(s,a)=\mathbb{E} \left[r^{(t+1)} \middle|s^{(t)}=s,a^{(t)}=a \right]$ is the expected reward of taking action $a$ at state $s$, and $\mathcal{P}^{a}_{ss'} = \Pr\left(s^{(t+1)}=s'\middle| s^{(t)} =s, a^{(t)}=a\right)$ is the transition probability from state $s$ to next state $s'$ with action $a$. The classical Q-learning algorithm uses a lookup table to represent the Q-function values and employs the fixed-point relation in \eqref{eq:bellmaneq} to iteratively update these values. However, the classical lookup table approach is not practical for continuous or large discrete state spaces. 
	
	To overcome this drawback, deep Q-learning replaces the lookup table with a deep neural network which is called deep Q-network and expressed as $q(s,a;\bm{\psi})$ with $\bm{\psi}$ being its parameters \cite{mnih2015human}. As described in \cite[Fig. 1]{mnih2015human}, its input layer is fed by a given state $s$, and each port of its output layer gives the Q-function value for input $s$ and corresponding action output. Deep Q-learning is an off-policy learning method that stores the past experiences in an experience replay memory denoted as $\mathcal{D}$ in the form of $e=(s,a,r',s')$. A small value for the maximum size of this memory, $|\mathcal{D}|$, will result with over-fitting, while a large value will slow down learning. Additionally, deep Q-learning adopts ``quasi-static target network'' technique that implies creating a target network with parameters $\bm{\psi}_\textrm{target}$ to predict the target values in the following mean-squared Bellman error:
	\begin{align}\label{eq:DQNMSBEloss}
	L\left(\bm{\psi},\mathcal{D}\right) &= \mathbb{E}_{(s,a,r',s') \sim \mathcal{D}} \left[\left( y(r',s') -q\left(s,a;\bm{\psi}\right)\right)^2\right],
	\end{align}
	where the target $y(r',s') = r' + \gamma \max_{a'}q\left(s',a';{\bm{\psi}_{\textrm{target}}}\right)$. To minimize \eqref{eq:DQNMSBEloss}, $\bm{\psi}$ is updated by sampling a random mini-batch $\mathcal{B}$ from $\mathcal{D}$ and running gradient descent by 
	\begin{align}\label{eq:criticgradient}
	\nabla_{\bm{\psi}} \frac{1}{|\mathcal{B}|} \sum_{(s,a,r',s') \in \mathcal{B}} \left(y(r',s') - q\left(s,a;\bm{\psi}\right) \right)^2.
	\end{align}
	Each iteration is followed by updating $\bm{\psi}_\textrm{train}$ by $\bm{\psi}$. During the training, instead of fully exploiting the updated policy, the learning agent applies the $\epsilon$-greedy strategy which takes a random action with a probability of $\epsilon$ for exploration.
	
	On the other hand, to overcome the challenge of applying deep Q-learning to continuous action spaces, Reference \cite{lillicrap2015ddpg} had proposed an actor-critic learning scheme called deep deterministic policy gradient. It iteratively trains a critic network, defined by $\phi$, to represent an action-value function, and uses the critic network to train an actor network, defined by $\bm{\theta}$, that parameterizes a deterministic policy. We define the deterministic policy as $\mu: \mathcal{S} \to \mathcal{A}$, and for a given state $s$, the action is determined by $a=\mu(s;\bm{\theta})$. Hence, the target policy $\mu^*$ satisfies the Bellman property:
	\begin{align}\label{eq:actorbellmaneq}
	Q^{\mu^*}(s,a)=\mathcal{R}(s,a)+\gamma \sum_{s' \in S}\mathcal{P}^{a}_{ss'}Q^{\mu^*}(s',\mu^*(s')),
	\end{align}
	
	Similar to deep Q-learning, the critic network is trained by minimizing the mean-squared Bellman error defined in \eqref{eq:DQNMSBEloss}. However, compared to the deep Q-network, the critic network has only one output that gives a Q-function value estimate for a given state and action input. In addition, the target in \eqref{eq:DQNMSBEloss} becomes $y_\textrm{critic}(r',s') = r' + \gamma q\left(s',\mu(s';\bm{\theta});{\bm{\phi}_{\textrm{target}}}\right)$. 
	
	Since $q(s,a;\bm{\phi})$ is differentiable with respect to action, caused by action space being continuous, the policy parameters are simply updated by the following gradient:
	\begin{align}\label{eq:actorgradient}
	\nabla_{\bm{\theta}} \frac{1}{|\mathcal{B}|} \sum_{(s,\dots) \in \mathcal{B}} q\left(s,\mu(s;\bm{\theta});\bm{\phi}\right).
	\end{align}
	Note that a noise term is added to the deterministic policy output for exploration during training.
	\begin{figure*}
		[t!]
		\centering
		\includegraphics[clip, trim=9.75cm 0.0cm 14.5cm 0.00cm,width=1.75\columnwidth]{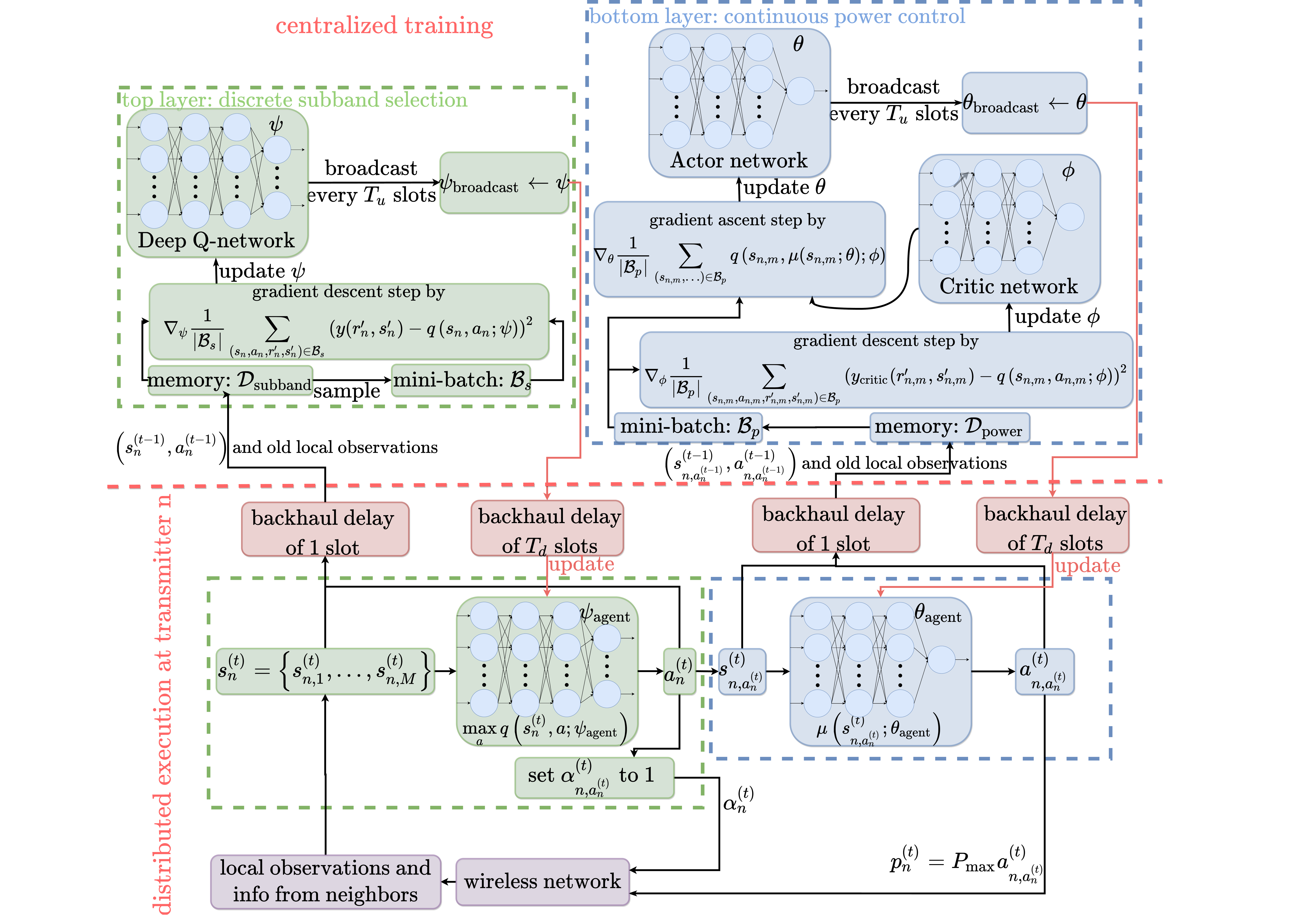}
		\caption{Diagram of the proposed power control algorithm.}
		\label{fig:crosslayerdiagram}
	\end{figure*}
	\subsection{Local Information and Neighborhood Sets}
	We next describe the extent of the local information at transmitter $n$ at the beginning of time slot $t$. At time $t$, transmitter $n$ has two types of neighborhood sets for each subband. The first set is called ``interferers'' that consists of $c$ indexes and is denoted as $\mathcal{I}^{(t)}_{n,m}$. For subband $m$, transmitter $n$ first divides nearby transmitters into two groups whether they used subband $m$ during time slot $t-1$ or not in order to prioritize the transmitters that occupy subband $m$. Then, it sorts each group according to the interfering channel strength at receiver $n$ from their transmitters during time slot $t-1$ by descending order, i.e., $g^{(t-1)}_{i \to n, m}$. Lastly, the first $c$ sorted nearby transmitters forms $\mathcal{I}^{(t)}_{n,m}$. 
	
	The second set is the set of ``interfered receivers'' that consists of $c$ indexes and is defined as $\mathcal{O}^{(t)}_{n,m}$. Again, each nearby receiver $j$ is first divided into two groups based on $\alpha^{(t-1)}_{j,m}$. The sorting criteria within each group becomes the potential significance of the interference strength at receiver $j$ from transmitter $n$ during time slot $t-1$, i.e., $g^{(t-1)}_{n \to j, m}\left(\sum_{l\in\mathcal{N},l\neq j}\alpha^{(t-1)}_{l,m}g^{(t-1)}_{l \to j, m}p^{(t-1)}_l+\sigma^2\right)^{-1}$. %$\alpha^{(t-1)}_{n,m}g^{(t-1)}_{n \to j, m}p^{(t-1)}_n \left(\sum_{l\in\mathcal{N},l\neq j}\alpha^{(t-1)}_{l,m}g^{(t-1)}_{l \to j, m}p^{(t-1)}_l+\sigma^2\right)^{-1}$
	
	Compared to \cite{nasir2018deep}, we follow simpler practical constraints on the available local information to be used in the state set design, as our main goal is to show the usefulness of the proposed approach. At the beginning of time slot $t$, transmitter $n$ has access to the most recent local information gathered at receiver $n$ for each subband $m$ such as $g^{(t)}_{n \to n, m}$, $g^{(t)}_{i \to n, m}$ $\forall{i\in \mathcal{I}^{(t)}_{n,m}}$, and sum interference power at receiver $n$, i.e., $\sum_{l\in\mathcal{N},l\neq n}\alpha^{(t-1)}_{l,m}g^{(t)}_{l \to n, m}p^{(t-1)}_l$. Conversely, the channel measurements gathered at nearby receivers are delayed by one time slots, e.g., $g^{(t-1)}_{n \to j, m}$ $\forall{j\in \mathcal{O}^{(t)}_{n,m}}$. Apart from the channel measurements, we assume that each interfered and interferer neighbor also sends crucial key performance indicators delayed by one time slot due to network latency, e.g., its achieved spectral efficiency during last slot.
	\subsection{Proposed Multi-Agent Learning Scheme}
	In order to allow distributed execution, each link, specifically, each transmitter, operates as an independent learning agent by treating other agents as part of its local environment. Hence, our approach is based on multiple learning agents, rather than a single learning agent that controls the entire action space whose dimensions will grow exponentially with the total number of links. The single learning agent approach has similar drawbacks as the conventional centralized optimization algorithms in terms of complexity and cost of communication. In contrast, the proposed multi-agent approach is easily scalable to larger networks and can operate with just local information after training.

	At the beginning of each time slot, each agent successively executes two policies to determine its associated subband and transmit power level. The reinforcement learning component at the top layer is a deep Q-network that is responsible for the subband selection. The bottom layer uses deep deterministic policy gradient algorithm to train the actor network responsible for agent's transmit power level decisions. As described in Fig. \ref{fig:crosslayerdiagram}, the actor network at the bottom layer requires the subband decision of the top layer to determine its state input before setting agent's transmit power.
	
	We next describe key components of the proposed design:
	\begin {enumerate}[leftmargin=*]
	\item\textbf{Action Set Design:} All agents have the same pair of action spaces. The top layer uses a discrete action space that consists of subband indexes, i.e, $a_n^{(t)} \in \mathcal{A}_\textrm{subband}=\left\{ 1,\dots,M\right\}=\mathcal{M}$. Hence, we denote the subband selection of agent $n$ for time slot $t$ as $a_n^{(t)}$. The bottom layer has a continuous action space defined as $\mathcal{A}_\textrm{power}= \left[0,1\right]$. Since the bottom layer is executed after the top layer, we denote its action as $a_{n,a^{(t)}_n}^{(t)}$. We later multiply it by $P_\textrm{max}$ to get $p_n^{(t)}=P_\textrm{max}a_{n,a^{(t)}_n}^{(t)}$.
	
	\item\textbf{State Set Design:} To be used in the state, all agents rank the subbands at the beginning of each time slot according to their direct channel gain to the total interference power ratio. We denote the rank as $z_{n,m}^{(t)}$. Now we describe the state of agent $n$ on subband $m$ at time $t$ as:

	\begin{align}\label{eq:state}
	\begin{split}
	s_{n,m}^{(t)} &= \Biggl\{\alpha_{n,m}^{(t-1)}p_n^{(t-1)}, C_{n}^{(t-1)}, z_{n,m}^{(t)}, g_{n \to n, m}^{(t)},\\& \sum_{l \neq n}\alpha^{(t-1)}_{l,m} g^{(t)}_{l \to n, m} p^{(t-1)}_{l}, \Bigl\{ g^{(t)}_{{i}\to n, m},\alpha_{{i},m}^{(t-1)} p_{i}^{(t-1)}, \\&  C_{i}^{(t-1)}, z_{i,m}^{(t-1)} \Bigl|\forall i \in \mathcal{I}^{(t)}_{n,m} \Bigr\},\Bigl\{g^{(t-1)}_{n\to {j}, m},g^{(t-1)}_{j\to j, m},\\ 
	 &C_{{j}}^{(t-1)}, z_{{j,m}}^{(t-1)},\sum_{l\neq j}\alpha^{(t-1)}_{l,m}g^{(t-1)}_{l \to j, m}p^{(t-1)}_l \Bigl| \forall j \in \mathcal{O}^{(t)}_{n,m}\Bigr\} \Biggr\}.
	\end{split}
	\end{align}
	
	Since the top layer does the subband decisions that requires information from all subbands, it should have a broader environment view than the bottom layer. Thus, for the top layer, we define agent $n$'s state as $s_n^{(t)}=\left\{s_{n,1}^{(t)},\dots,s_{n,M}^{(t)}\right\}$. Then, the bottom layer uses $s_{n,a^{(t)}_n}^{(t)}$ as its input. 
	\item\textbf{Reward Function Design:} Both learning layers collaboratively aim to maximize the objective in \eqref{eq:objective}. Consequently, they share the same reward function that describes the overall contribution of agent's combined subband and power decisions on the sum-rate objective. This includes agent's own spectral efficiency and a penalty term depending on its externalities to its interfered neighbors on subband $a_n^{(t)}$ \cite{nasir2018deep}. For the reward function, we first compute the externality of agent $n$ to interfered $j \in \mathcal{O}^{(t+1)}_{n,a^{(t)}_n}$ during time slot $t$ as
	\begin{align}\label{eq:externality}
	\begin{split}
	\pi^{(t)}_{n\rightarrow j} &= C^{(t)}_{j \backslash n,a^{(t)}_n} - C^{(t)}_{j,a^{(t)}_n},
	\end{split}
	\end{align}
	where $C^{(t)}_{j \backslash n,a^{(t)}_n}$ is the spectral efficiency of $j$ without the interference from agent $n$ on subband $a^{(t)}_n$ during slot $t$:
	\begin{align}\label{eq:spectralwithoutn}
	\begin{split}
	C^{(t)}_{j \backslash n,a^{(t)}_n} &= \log \left(1+\frac{\alpha^{(t)}_{j,a^{(t)}_n} g^{(t)}_{j \to j, a^{(t)}_n} p^{(t)}_{j}}{\sum_{l \neq n,j}\alpha^{(t)}_{l,a^{(t)}_n} g^{(t)}_{l \to j, a^{(t)}_n} p^{(t)}_{l}+\sigma^2} \right).
	\end{split}
	\end{align}
	
	Next, we define the reward of agent $n$ as
	\begin{align}\label{eq:reward}
	\begin{split}
	r^{(t+1)}_n &= C^{(t)}_{n,a^{(t)}_n} - \sum_{j \in \mathcal{O}^{(t+1)}_{n,a^{(t)}_n}} \pi^{(t)}_{n\rightarrow j}.
	\end{split}
	\end{align}
	
	\item\textbf{Centralized Training:} Since multi-agent setting violates the environment stationary assumption of the underlying Markov decision process discussed in Section \ref{sec:RLdiscussion}, there is an extensive research to develop multi-agent learning frameworks with good empirical performance, but rarely with theoretical guarantees\cite{nguyen2018multisurvey}. In this work, we ensure the stability by training global policy parameters shared across the network and trained by a centralized trainer that gathers experiences of all agents. As shown in Fig. \ref{fig:crosslayerdiagram}, centralized training stores two experience-replay memories for each layer: $\mathcal{D}_\textrm{subband}$ and $\mathcal{D}_\textrm{power}$. At time $t$, the most recent experience at $\mathcal{D}_\textrm{subband}$ and $\mathcal{D}_\textrm{power}$ from agent $n$ is $e_{n,\textrm{subband}}^{(t-1)}=\left(s_n^{(t-2)},a_n^{(t-2)},r_n^{(t-1)},s_n^{(t-1)}\right)$ and $e_{n,\textrm{power}}^{(t-1)}=\left(s_{n,a_n^{(t-2)}}^{(t-2)},a_{n,a_n^{(t-2)}}^{(t-2)},r_n^{(t-1)},s_{n,a_n^{(t-2)}}^{(t-1)}\right)$, respectively, due to the backhaul delay of 1 time slot. Note that the next state in $e_{n,\textrm{power}}^{(t-1)}$ is with respect to the old subband selection $a_n^{(t-2)}$.
	
	During time slot $t$, the centralized training runs one gradient step for each policy. As described in Fig \ref{fig:crosslayerdiagram}, it broadcasts most recent versions of $\psi$ and $\theta$ once per $T_u$ time slots. The broadcasting takes $T_d$ time slots to finish, again due to the backhaul delay. 
	\end {enumerate}
	\section{Simulation Results}\label{sec:results}
	In this section, our main goal is to compare the performance of the proposed learning approach with the conventional optimization methods and joint learning as the number of subbands increases.% First, we describe the channel model and some specifics about the centralized training, and then present the results on various test scenarios.
	
	\begin{figure}
		[t]
		\centering
		\includegraphics[clip, trim=0.00cm 0.0cm 0.0cm 0.00cm,width=1.0\columnwidth]{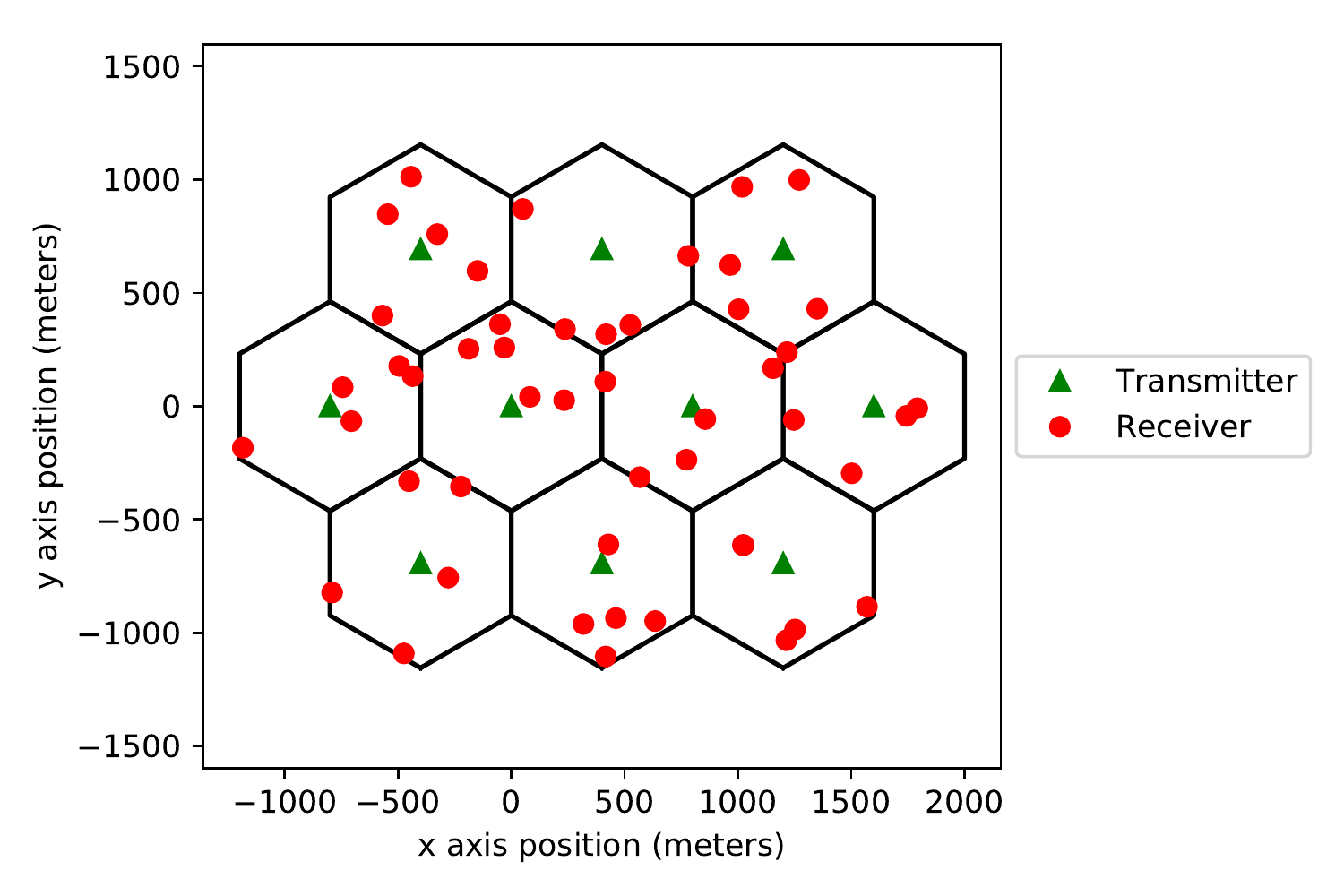}
		\caption{A network configuration example.}
		\label{fig:deployment}
	\end{figure}
	\begin{table*}[t]	
	\footnotesize
	\tabcolsep 0pt \caption{Testing results.}
	\begin{center} 
		\def\temptablewidth{2\columnwidth}
		{\rule{\temptablewidth}{1pt}}
		\begin{tabular*}{\temptablewidth}{@{\extracolsep{\fill}}|c|c|cc|ccc|cc|c|}
			{}&{}&\multicolumn{5}{c|}{average sum-rate performance in bps/Hz per link}&\multicolumn{2}{c|}{output layer size}&{average}\\
			{$(K,N)$} &{$M$} &  \multicolumn{2}{c|}{reinforcement learning} & \multicolumn{3}{c|}{other schemes}&\multicolumn{2}{c|}{reinforcement learning} & {iterations} \\
			{(cells, links)} &{subbands} &proposed &joint & ideal FP & delayed FP & random & proposed &joint & FP
			\\\hline \hline
			{} 			& 1 & 1.51 & 1.50 & 1.58 & 1.46 & 0.41 & {1 \texttt{+} 1} & {10} & 70.30 \\
			{$(5,20)$} 	& 2 & 2.63 & 2.64 & 2.66 & 2.46 & 0.99 & {2 \texttt{+} 1} & {20} & 102.08 \\
			{} 			& 4 & 4.57 & 4.38 & 3.81 & 3.57 & 2.12 & {4 \texttt{+} 1} & {40} & 122.15 \\ \hline \hline
			{} 			& 1 & 1.26 & 1.26 & 1.31 & 1.21 & 0.25 & {1 \texttt{+} 1} & {10} & 72.83 \\
			{} 			& 2 & 2.08 & 2.10 & 2.08 & 1.92 & 0.59 & {2 \texttt{+} 1} & {20} & 96.32 \\
			{$(10,50)$}  & 4 & 3.34 & 3.34 & 2.90 & 2.68 & 1.31 & {4 \texttt{+} 1} & {40} & 185.93 \\
			{} 			& 5 & 3.79 & 3.76 & 3.18 & 2.94 & 1.64 & {5 \texttt{+} 1} & {50} & 206.38 \\
			{} 			& 10 & 5.71 & 4.41 & 4.44 & 4.08 & 2.99 & {10 \texttt{+} 1} & {100}& 287.70
		\end{tabular*}
		{\rule{\temptablewidth}{1pt}}
	\end{center}
	\label{table:testing_smallnetwork}
\end{table*}
	\begin{figure*}
		[t]
		\centering
		\subfloat[$M=2$ subbands, $(K,N)=(5,20)$.]{
			\includegraphics[width=0.935\columnwidth]{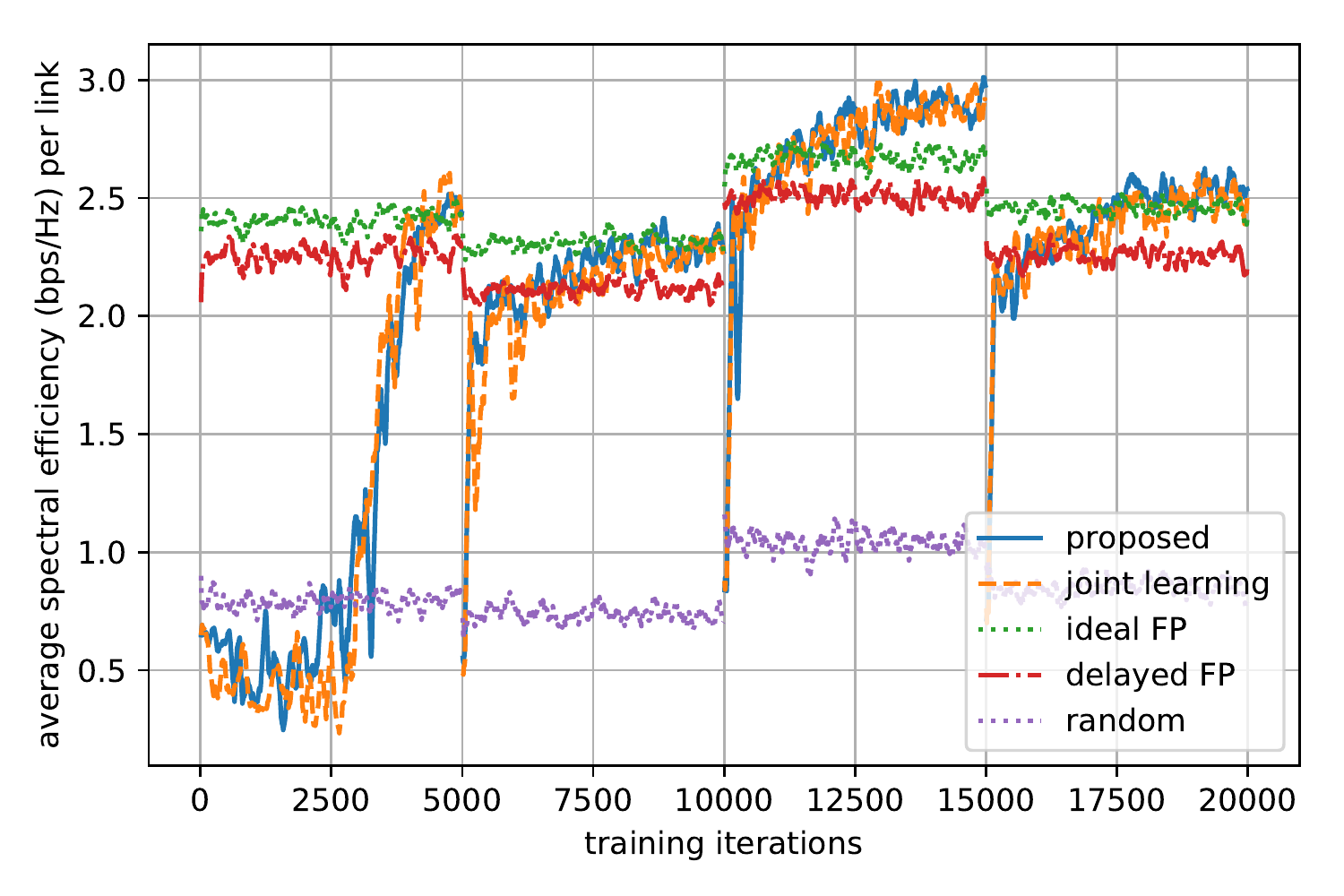}
			\label{fig:training1}}
		\hfil
		\subfloat[$M=4$ subbands, $(K,N)=(5,20)$.]{
			\includegraphics[width=0.935\columnwidth]{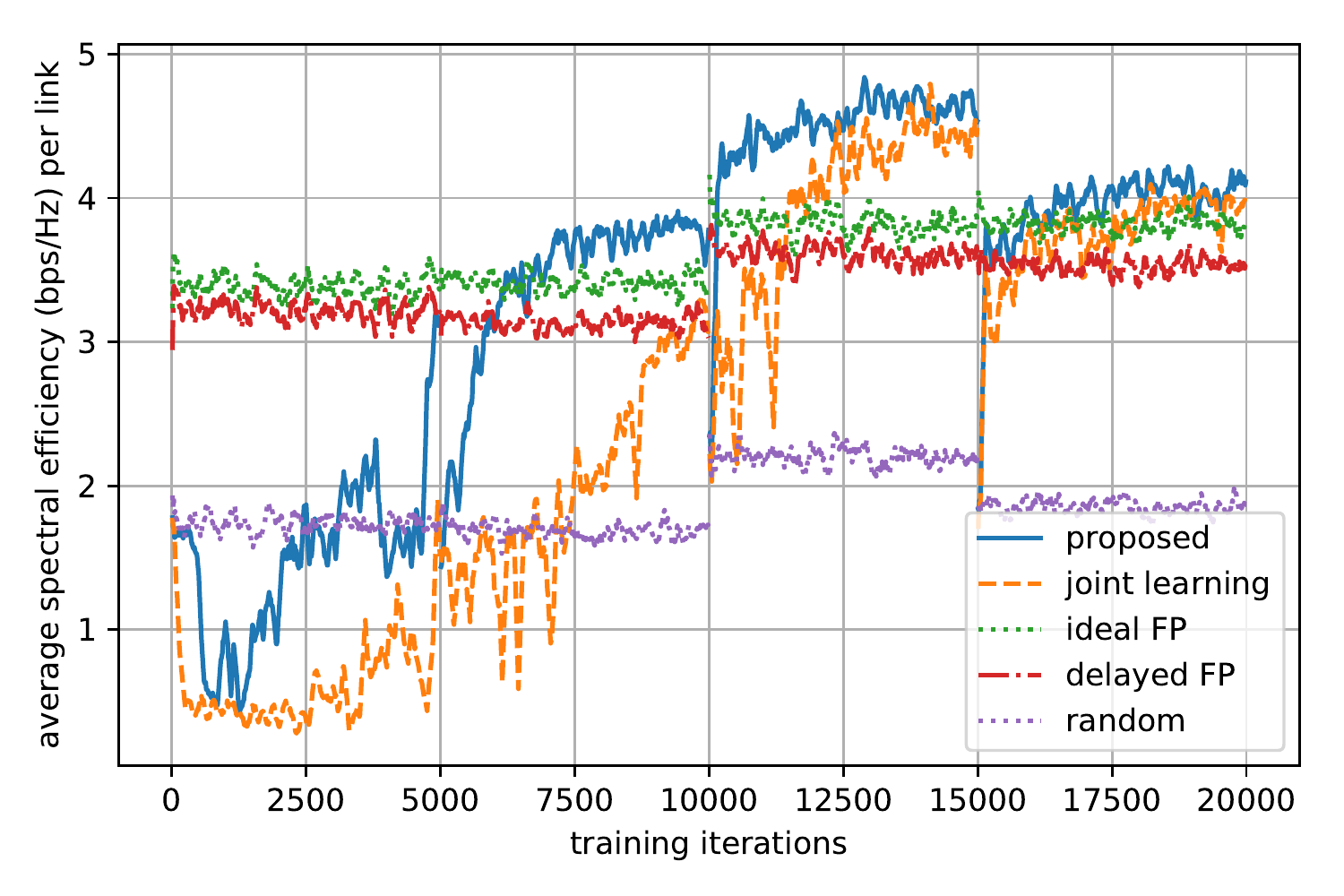}
			\label{fig:training2}}
		\hfil
		\subfloat[$M=5$ subbands, $(K,N)=(10,50)$.]{
			\includegraphics[width=0.935\columnwidth]{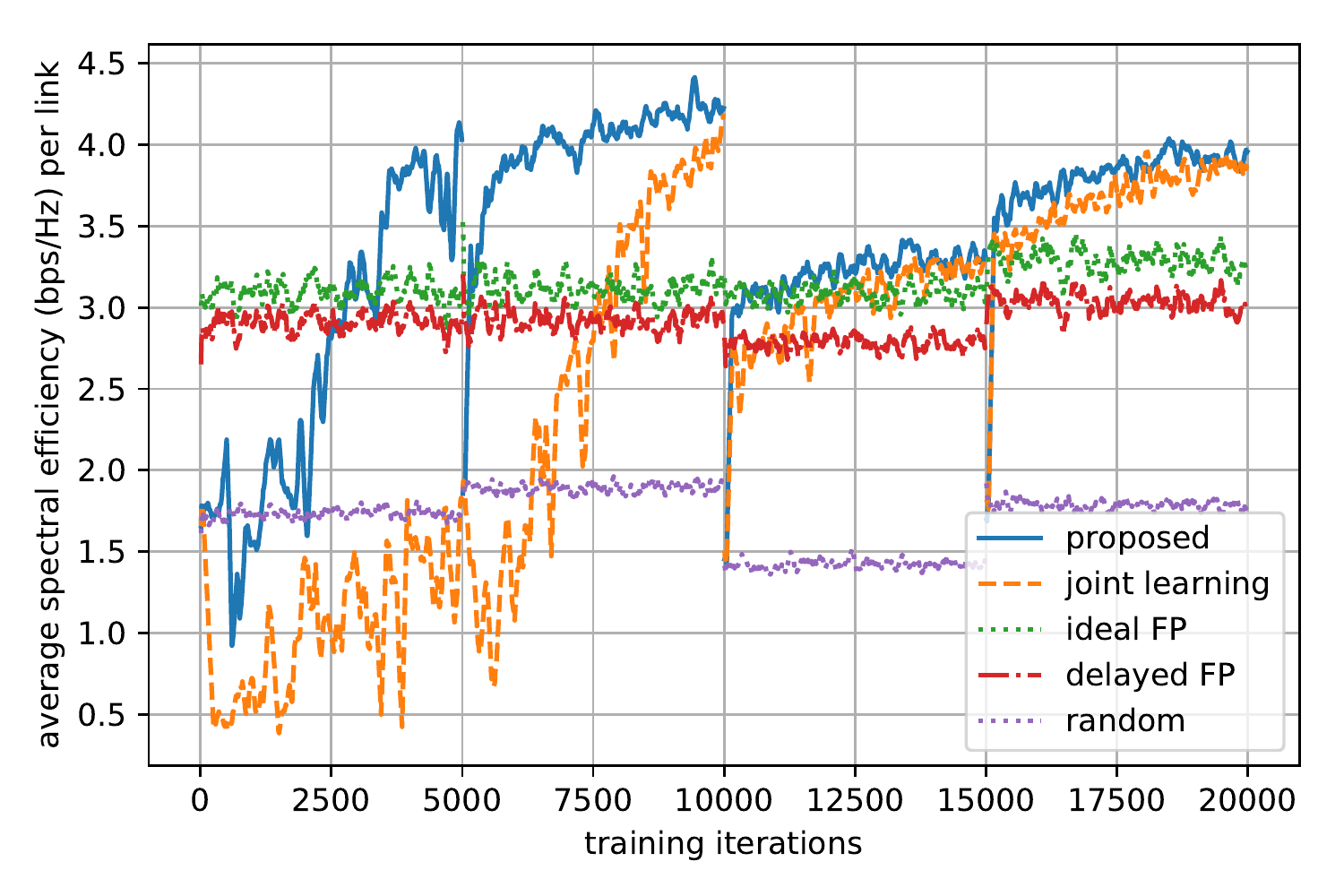}
			\label{fig:training3}}
		\hfil
		\subfloat[$M=10$ subbands, $(K,N)=(10,50)$.]{
			\includegraphics[width=0.935\columnwidth]{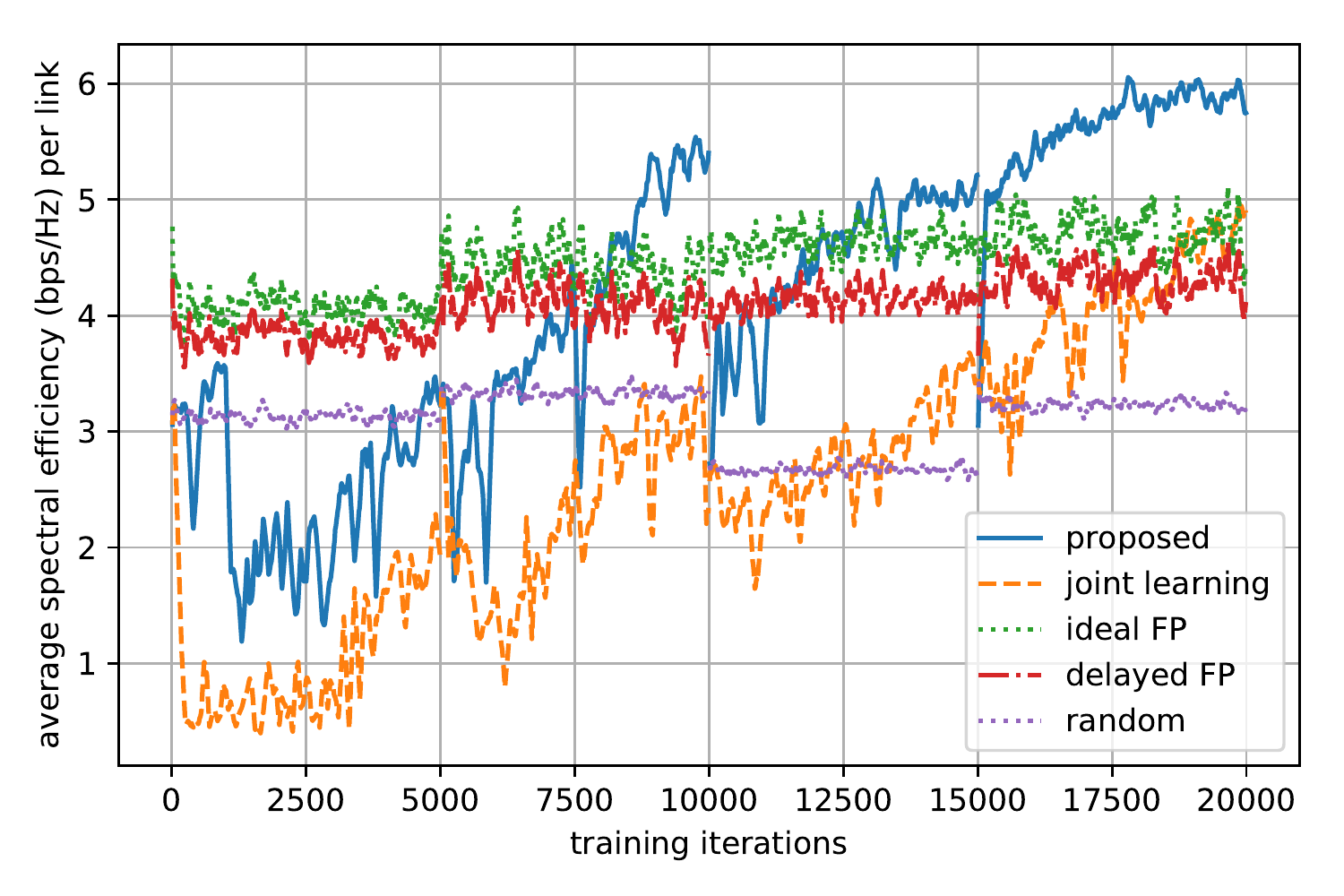}
			\label{fig:training4}}
		\caption{Training convergence.}
		\label{fig:training}
	\end{figure*}
	Throughout the simulations, we choose two network sizes of $(K,N)=(5\textrm{ cells},20\textrm{ links})$ and $(10\textrm{ cells},50\textrm{ links})$, respectively. As described in Fig. \ref{fig:deployment}, we consider homogeneous hexagonal cells of 400 meters radius with each cell having equal number of uniformly randomly placed receivers. We vary the number of subbands $M$ from $1$ to $10$. Following the LTE standard, we set the distance dependent path loss to $128.1 + 37.6\log_{10}(d)$ (in dB), where $d$ is transmitter-to-receiver distance in km. The log-normal shadowing standard deviation is $10$ dB. We set $f_d = 10$ Hz, $T = 20$ ms, $P_{\textrm{max}}= 38$ dBm, and $\sigma^2=-114$ dBm. Similar to \cite{nasir2018deep}, the signal-to-interference-plus-noise ratio is capped at $30$ dB in the calculation of the spectral efficiency in \eqref{eq:DynRate} due to practical constraints on front end's dynamic range.
		
	We compare the proposed approach with four benchmarks. The first is the joint learning approach as proposed in \cite{tan2019jointDRL}. %We use a deep Q-network of three fully-connected hidden layers with 200, 200, 100 neurons, respectively. 
	We discretize the transmit power into $10$ levels. The second is called the `ideal FP'. It runs the fractional programming algorithm with an assumption of full instant CSI. The first scenario ignores any delay during the execution of centralized optimization or passing the optimization outcomes to the transmitters. On the other hand, the third benchmark is called the `delayed FP' and assumes one time slot delay to run the fractional programming algorithm. In the final benchmark, each transmitter just picks a random subband and transmit power at the beginning of every time slot.
	
%	\begin{figure}
%	[t]
%	\centering
%	\includegraphics[clip, trim=0.40cm 0.3cm 0.40cm 0.40cm,width=1.0\columnwidth]{exhaustive_vs_FP.pdf}
%	\caption{Comparison between fractional programming and an exhaustive search method on channel selection.}
%	\label{fig:exhaustive_vs_FP}
%	\end{figure}
%	To show the effect of multiple subbands on the suboptimal fractional programming approach which performs quite well for the single-band case, we compare the sum-rate performance of the fractional programming and an exhaustive search scheme to look for optimal channel selections. In Fig. \eqref{fig:exhaustive_vs_FP}, we vary the number of subbands for $200$ random tests by fixing $K$ to 5 cells and $N$ to 5 links. Note that the exhaustive search method runs the fractional programming independently on each subband and examines all $M^N$ possible subband permutations. Fig \ref{fig:exhaustive_vs_FP} shows that there is a quite large margin between exhaustive search and fractional programming which is a motivation for a distributed deep reinforcement approach that can match or outperform the fractional programming.
	We divide training into four episodes with each running for 5,000 time slots. At the beginning of each episode, we randomly sample a new deployment, and we reset the exploration and learning rate parameters. For faster convergence, we replace the noise term added to the deterministic policy output with Q-learning's $e$-greedy algorithm. The implementation and hyper-parameters are included in the source code which is available at \cite{nasir2020githubJointPowerSubband}. For better stability, we ensure that the bottom layer has higher learning rate than the top layer, and it uses a higher initial value of $\epsilon$, but with a higher decay rate. The fine-tuning of the $\epsilon$ value is important to avoid converging to undesired situations in which all agents want to transmit with $P_\text{max}$ or with zero power.
		
	In Fig. \ref{fig:training}, we show the training convergence of the proposed and joint reinforcement learning scheme. For $M$ = 2 subbands, as shown in Fig. \ref{fig:training1}, their convergence rates are quite close. However, when we increase the number of subbands, the joint learning approach is not able to keep up with the proposed approach in terms of training convergence. This is mainly caused by the increased size of the joint learning's action space and increased deep Q-network output layer complexity. Next, we test the performance of the trained policies on several randomly generated deployments in Table \ref{table:testing_smallnetwork}. Testing shows that a pretrained policy is still usable on new deployments and the proposed approach is better scalable than the benchmarks. % In addition, compared to the joint learning and fractional programming, cross-layer learning's complexity scales better.	
	
	\section{Conclusion and Future Work}\label{sec:conclusion}
	We have demonstrated a novel multi-agent reinforcement learning framework for the joint subband selection and power control problem. With centralized training and distributed execution only local information is needed by the agent under practicality constraints. In addition, as the number of subbands increases, the proposed learning approach has better training convergence and higher sum-rate performance than the joint learning. For future work, we are looking into better and easily tunable training and exploration schemes to better adapt to the environment non-stationarity of the multi-agent setting.
	
	%\clearpage
	%\footnotesize{}
	\bibliographystyle{IEEEtran}
	\bibliography{ref}{}
\end{document}